\documentclass{moriond}
\newcommand{\Photo}{}

\def\be{\begin{equation}}
\def\ee{\end{equation}}
\def\bea{\begin{eqnarray}}
\def\eea{\end{eqnarray}}

\newcommand{\ms}{m_{s}}
\newcommand{\md}{m_{d}}

\begin{document}
\vspace*{4cm}
\title{CURRENT AND FUTURE CONSTRAINTS ON COSMOLOGY AND MODIFIED GRAVITATIONAL WAVE FRICTION FROM BINARY BLACK HOLES}

\author{K. LEYDE$^{1,\, }$\footnote{\href{kleyde@apc.in2p3.fr}{kleyde@apc.in2p3.fr}} \footnote{To appear in the proceedings of the 56th Rencontres de Moriond 2022}, S. MASTROGIOVANNI$^{1,2}$, D. A. STEER$^{1}$, E. CHASSANDE-MOTTIN$^{1}$, C. KARATHANASIS$^{3}$}

\address{$^{1}$Universit\'e Paris Cit\'e, CNRS, Astroparticule et Cosmologie, F-75006 Paris, France\\
$^{2}$Artemis, Universit\'e C\^ote d'Azur, Observatoire de la C\^ote d’Azur, CNRS, F-06304 Nice, France\\
$^{3}$Institut de F\'isica d'Altes Energies (IFAE), Barcelona Institute of Science and Technology, Barcelona, Spain}

\maketitle\abstracts{
In this proceedings, we are interested in \textit{dark gravitational wave standard sirens} and their use for cosmology and for constraining modified gravity theories. Due to the extra friction term introduced in their propagation equation those theories predict different luminosity distances for electromagnetic and gravitational waves (GWs). This effect can be parametrized by the two variables $\Xi_0$ and $n$, that can be measured from gravitational wave observations, and specifically from the binary black hole (BBH) mergers detected by LIGO and Virgo. By fitting jointly BBH population models in mass and redshift, the cosmological parameters, and the modified GW luminosity distance to $\sim$ 60 signals observed during the first three LIGO/Virgo observation runs, we conclude that general relativity is the preferred model. The future observation runs O4 and O5 are also considered. We forecast a measurement uncertainty on the modified gravity parameter $\Xi_0$ of $51\%$ with O4, and $20\%$ with O4 and O5, respectively, if GR is correct. However, we underline that there are strong correlations between astrophysical, cosmological and modified gravity parameters, possibly leading to bias if wrong priors are applied. A more detailed version of this communication can be found here.\cite{Leyde:2022orh}}
%
The use of \textit{dark gravitational wave (GW) standard sirens} in this work relies on the relation between the source frame mass $m_s$ and the detector frame (observed) mass $\md$\,, namely $\md = (1+z)\ms$. 
This relation, when combined with the assumption that BBHs stem from a unique and universal mass distribution, allows to infer the source redshift from the detector frame mass obtained from the GW data. We investigate if, in addition to populations and cosmology, dark sirens can also be used to test theories of modified gravity, and constrain deviations from GR at cosmological scales. We are specifically interested in beyond GR theories which modify the GW propagation equation.\cite{Belgacem:2017ihm}
%
%
It is assumed that these theories leave the evolution of the cosmic background unchanged, i.e. Friedmann's equations describe the expansion of the Universe ($H_0$ and $\Omega_{\rm m}$ for a flat $\Lambda$CDM model). 
We consider a phenomenological parametrization of the GW luminosity distance $d_L^{\rm GW}$ that is given by \cite{Belgacem:2017ihm}
\begin{equation}
\label{eq: def Xi parametrization dl}
    d_{L}^{\mathrm{GW}} = d_{L}^{\mathrm{EM}}\left(\Xi_0 + \frac{1-\Xi_0}{(1+z)^n}\right)\,.
\end{equation}
\section{Results with GWTC-3}
Using the most recent BBH catalog of the LIGO-Virgo-KAGRA collaboration, we perform a model selection analysis between GR vs the $\Xi_0$ gravity model, and between four different BBH mass distributions. We find that GR and the \textsc{multi peak} mass distribution are consistently the preferred models irrespective of the chosen selection cuts.
All measurements of $\Xi_0$ are consistent with the value predicted with GR ($90\%$ confidence level). We note that the $\Xi_0$ measurement depends on the assumptions made on the BBH mass model. This underlines the importance of checking the robustness of the conclusion using more than one BBH mass model. The modified gravity parameter $\Xi_0$ is strongly correlated with astrophysical and cosmological parameters such as the redshift evolution parameters and the Hubble constant. The results for $\Xi_0$ are compatible at the 1-$\sigma$ level with previous works with O3 data using the broken power law mass model.\cite{Mancarella:2021ecn}
\begin{figure}
\caption[]{The marginalized posteriors of the modified gravity parameter $\Xi_0$ of Eq.~\ref{eq: def Xi parametrization dl} that modifies the GW luminosity distance. \textit{(Left)} The $\Xi_0$ posterior from GWTC-3 data (the GR value is indicated as a dashed line). The three colors correspond to the SNR cuts of 10 (green), 11 (blue) and 12 (orange).
\textit{(Right)} Forecast for the $\Xi_0$ posterior with 510 events with an O4$+$O5 scenario, assuming wide priors on the cosmology (blue/\textit{Wide}), or Planck uncertainties (orange/\textit{Planck}). }
\vspace{0.3cm}
\begin{minipage}{0.48\linewidth}
\centerline{\includegraphics[width=0.70\linewidth]{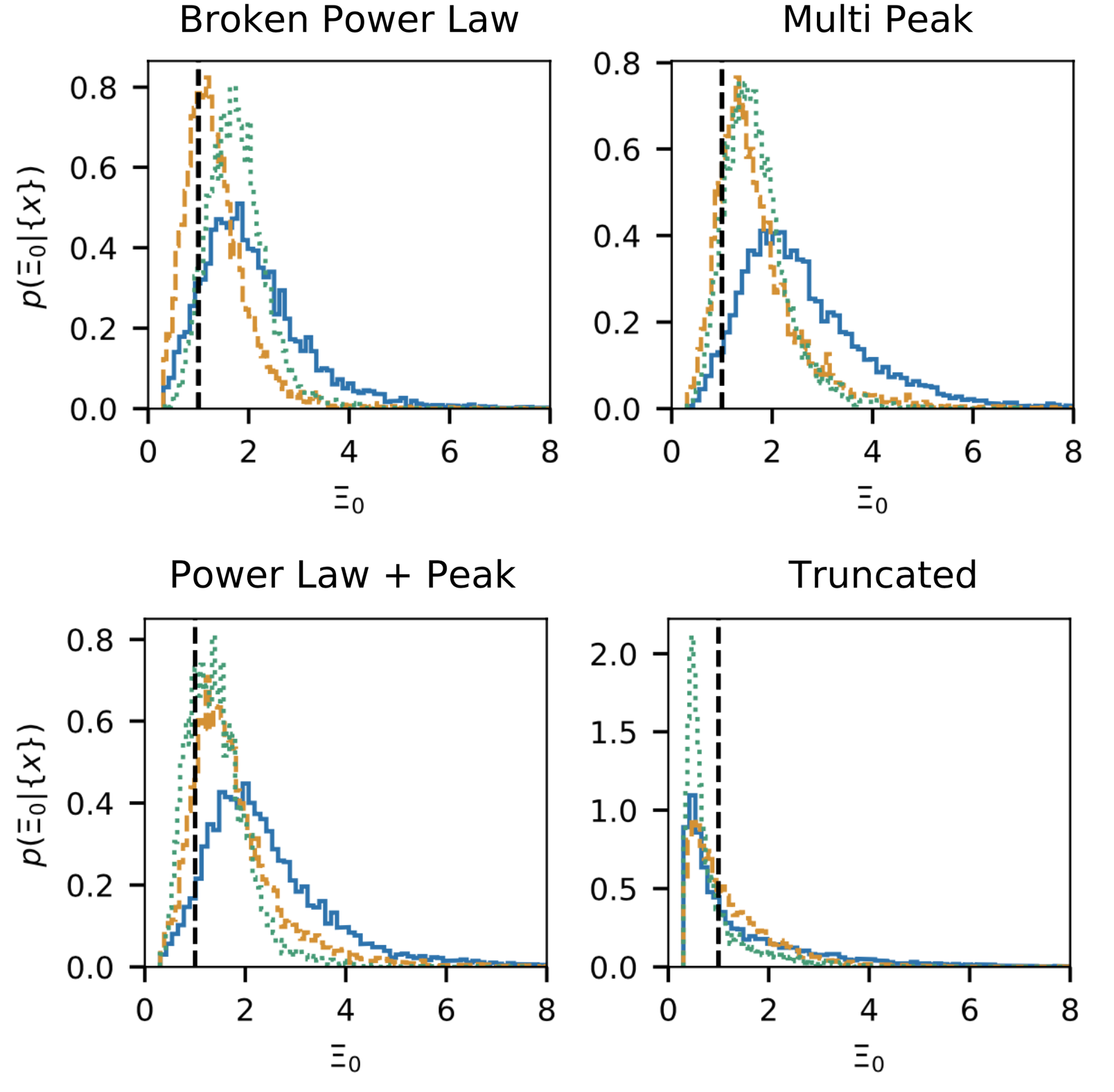}}
\end{minipage}
\hfill
\begin{minipage}{0.5\linewidth}
\centerline{\includegraphics[width=0.9\linewidth]{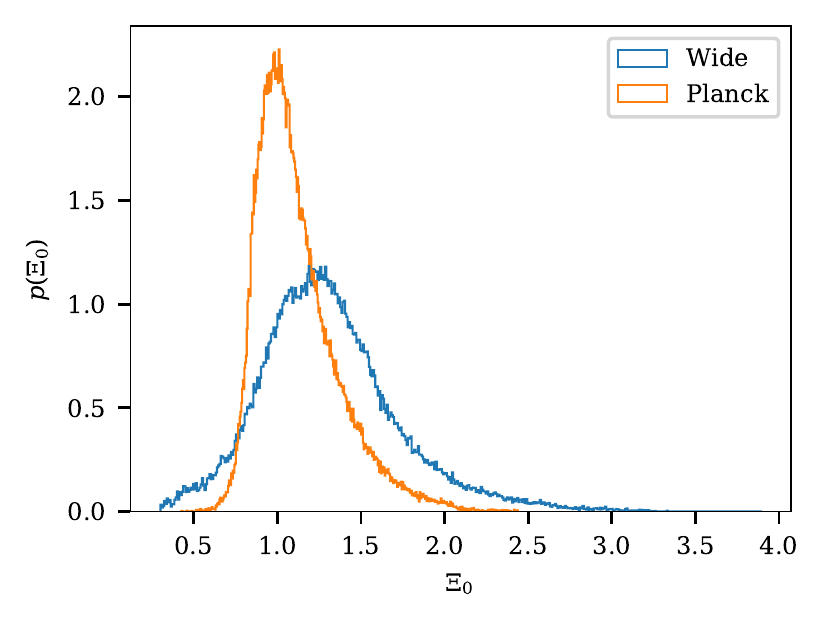}}
\end{minipage}
\label{figure1}
\end{figure}
\section{Forecast for O4 and O5}
We simulate the expected outcome of future O4 and O5 runs which results in 87 and 423 observed BBH events, resp. We find that if GR is the true theory of gravity, and assuming narrow priors on the cosmological parameters (from Planck), we recover the modified gravity parameter with a precision of $51\%$ with O4, and $20\%$ with O4 and O5 combined, cf.~Fig.~\ref{figure1} \textit{(right)}. The uncertainty on $\Xi_0$ is increased by a factor of 1.5 when using agnostic priors on $H_0$ and $\Omega_{\rm m}$.

\vspace{2mm}
{\footnotesize
\noindent \textbf{Acknowledgments} --- We thank the LIGO Scientific Collaboration and Virgo Collaboration for very helpful discussion and giving  access to the parameter estimation software used. The Fondation CFM pour la Recherche in Paris has generously supported KL during his doctorate.}
\section*{References}
\bibliography{references}

\end{document}